# From Glue-Code to Protocols: A Critical Analysis of A2A and MCP Integration for Scalable Agent Systems

Qiaomu Li
*College of Computing and Software Engineering*
*Kennesaw State University*
Marietta, GA 30060, USA
qli12@students.kennesaw.edu

Ying Xie
*College of Computing and Software Engineering*
*Kennesaw State University*
Marietta, GA 30060, USA
yxie2@kennesaw.edu

## Abstract

Artificial intelligence is rapidly evolving towards multi-agent systems where numerous AI agents collaborate and interact with external tools. Two key open standards, Google's Agent to Agent (A2A) protocol for inter-agent communication and Anthropic's Model Context Protocol (MCP) for standardized tool access, promise to overcome the limitations of fragmented, custom integration approaches. While their potential synergy is significant, this paper argues that **effectively integrating A2A and MCP presents unique, emergent challenges at their intersection, particularly concerning semantic interoperability between agent tasks and tool capabilities, the compounded security risks arising from combined discovery and execution, and the practical governance required for the envisioned "Agent Economy".** This work provides a critical analysis, moving beyond a survey to evaluate the practical implications and inherent difficulties of combining these horizontal and vertical integration standards. We examine the benefits (e.g., specialization, scalability) while critically assessing their dependencies and trade-offs in an integrated context. We identify key challenges increased by the integration, including novel security vulnerabilities, privacy complexities, debugging difficulties across protocols, and the need for robust semantic negotiation mechanisms. In summary, A2A+MCP offers a vital architectural foundation, but fully realizing its potential requires substantial advancements to manage the complexities of their combined operation.

## 1. Introduction and Motivation

The trajectory of Artificial Intelligence (AI) increasingly points towards complex ecosystems of interacting, specialized agents [24, 33]. These Multi-Agent Systems (MAS) promise enhanced capabilities but face fundamental challenges in coordination, communication, interaction with diverse external tools and data sources [15, 27], knowledge organization, and integrating expertise [43]. Historically, integrating heterogeneous agents and tools required bespoke, brittle "glue code," hindering the development of scalable and robust systems [10].

Two prominent open standards have emerged to address these bottlenecks: Anthropic's Model Context Protocol (MCP) standardizes the "vertical" interaction between an agent and its environment (tools, data), while Google's Agent-to-Agent (A2A) protocol standardizes the "horizontal" communication and collaboration between agents [1, 2, 5, 6]. MCP offers a universal interface for tool access, reducing the M x N integration problem for tools [42]. A2A provides a framework for agent discovery, task delegation, and secure inter-agent communication, offering a modern alternative to traditional, often semantically rigid, Agent Communication Languages (ACLs) [16]. Multi-agent architectures, leveraging specialized agents and coordination mechanisms, are being proposed as solutions to the limitations of standalone LLMs [17, 43].

While these protocols are explicitly designed to be complementary, their effective integration is not automatic. **This paper provides a critical analysis focused specifically on the synergy and difficulties arising from the *combination* of A2A and MCP.** Our central argument is that while integrating these standards offers substantial potential benefits, it also introduces unique, emergent challenges that must be addressed for realizing seamless, scalable agent systems. The novelty of our contribution lies in moving beyond surveying the individual protocols to critically evaluating the practical effects, architectural trade-offs, and specific risks *at the intersection* of horizontal (A2A) and vertical (MCP) integration.

We analyze how benefits like specialization are impacted by the need for semantic alignment between A2A tasks and MCP tools. We examine how security risks grow when A2A's agent discovery is combined with MCP's tool execution capabilities. We assess the feasibility of the "Agent Economy" from A2A/MCP interactions.

This analysis proceeds as follows: Section 2 recaps the core components of A2A and MCP. Section 3 presents a taxonomy of integration benefits, critically evaluating their practical realization *in an integrated context*. Section 4 discusses architectural integration patterns, highlighting potential difficulties *at the interface*. Section 5 analyzes the economic implications and the "Agent Economy" concept through a critical perspective. Section 6 delves into the risks and challenges specifically *increased or created* by the integration. Section 7 outlines future research directions to overcome these integration-specific hurdles. Figure 1 provides a conceptual overview.

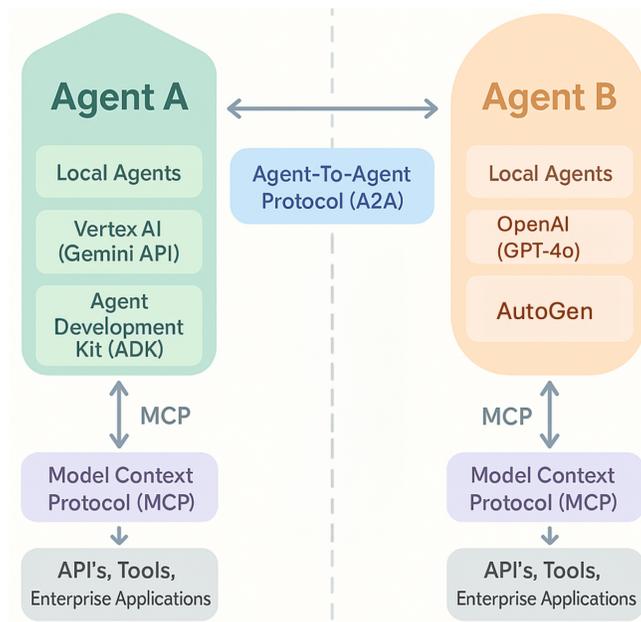

*Fig. 1.* *The multi-agent workflow with A2A + MCP*

## 2. Background: Protocol Components

The emergence of LLM-based agents has driven the need for standardized interaction mechanisms like A2A and MCP. Equipping LLMs with external tools is now a key paradigm [34, 40, 41].

### 2.1 Google's A2A Protocol: Enabling Agent Collaboration

The Agent-to-Agent (A2A) protocol is an open standard introduced by Google to enable direct communication and interoperability between AI agents across frameworks and vendors [1]. Its core purpose is to let agents coordinate actions and share information in a structured way, rather than treating other agents merely as tools or static API endpoints. A2A defines a client–server style interaction between agents: a **client agent** formulates tasks and requests, and a **remote agent** acts on those tasks, returning results or performing actions. Importantly, any agent can play either role depending on the context. Key components and features of A2A include:

- **Core Design Principles:** A2A is built on principles like embracing agentic capabilities (recognizing agents as autonomous entities), leveraging existing web standards (HTTP, JSON-RPC 2.0, SSE) for easier adoption, being secure by default (enterprise-grade authentication/authorization), supporting long-running asynchronous tasks, and being modality-agnostic (handling text, files, streams, JSON, etc.). The concept of "opaque execution" allows agents to collaborate based on advertised capabilities without needing internal implementation details.
- **Agent Discovery (Agent Cards):** A2A provides a mechanism for agents to advertise their capabilities in a standardized format called an **Agent Card**. Typically hosted at a `/.well-`

known/agent.json URL, this JSON document describes the agent's name, description, endpoint URL, provider info, version, authentication requirements (mirroring OpenAPI schemes), supported modalities (MIME types), and specific **skills** [1, 6]. Each skill includes an ID, name, description, tags, optional examples, and skill-specific input/output modes. This allows a client agent to discover which remote agent is best suited for a given task, potentially querying registries or marketplaces. Ensuring the security and reliability of this discovery process is itself a challenge [21, 28].

- **Tasks and Task Lifecycle:** The fundamental unit of work is a **Task** object, initiated by the client (e.g., via tasks/send) [1, 6]. Each task has a unique ID and encapsulates a request (including parameters, desired skill, etc.). It progresses through a defined lifecycle managed by the remote agent, with states like submitted, working, input-required, completed, failed, canceled, and unknown. Tasks facilitate the exchange of Messages and the generation of Artifacts. A Task object contains fields like id, skillId, state, optional sessionId, the initial message, and optional notification configurations.

- **Messaging and Parts:** Communications within a task occur via **Message** objects. Each message has a role (user or agent) and contains one or more **Part** objects representing the content [1, 6]. Parts are the fundamental content units and can be of different MIME types, supporting rich, multimodal communication (e.g., TextPart, FilePart for binary data/URIs, DataPart for structured JSON).

- **Artifacts:** Immutable outputs or results generated by the remote agent during task execution are represented as **Artifact** objects [1, 6]. A task can produce multiple artifacts (e.g., code and documentation). Artifacts contain Part objects and include fields like name, description, parts, metadata, and index for ordering. Optional append and lastChunk flags support streaming updates.

- **Communication Protocol (JSON-RPC & SSE):** A2A uses JSON-RPC 2.0 over HTTP(S) for remote procedure calls. Key methods include tasks/send (initiate task), tasks/sendSubscribe (initiate and subscribe via SSE), tasks/get (retrieve task status), tasks/cancel (request cancellation), and methods for managing push notifications (tasks/pushNotification/set, tasks/pushNotification/get).

- **Long-Running Tasks & Streaming Updates:** A2A explicitly supports long-running processes via asynchronous handling. Clients typically use tasks/sendSubscribe. The server responds immediately and establishes a persistent Server-Sent Events (SSE) connection (Content-Type: text/event-stream) to push updates (e.g., TaskStatusUpdateEvent, TaskArtifactUpdateEvent) as they occur. Webhook-based push notifications are also supported for disconnected clients.

- **UX Negotiation:** Agents can advertise supported input/output modes in their Agent Cards and use multi-part messages to offer content in various formats, allowing the client to select the best representation for its UI or capabilities.

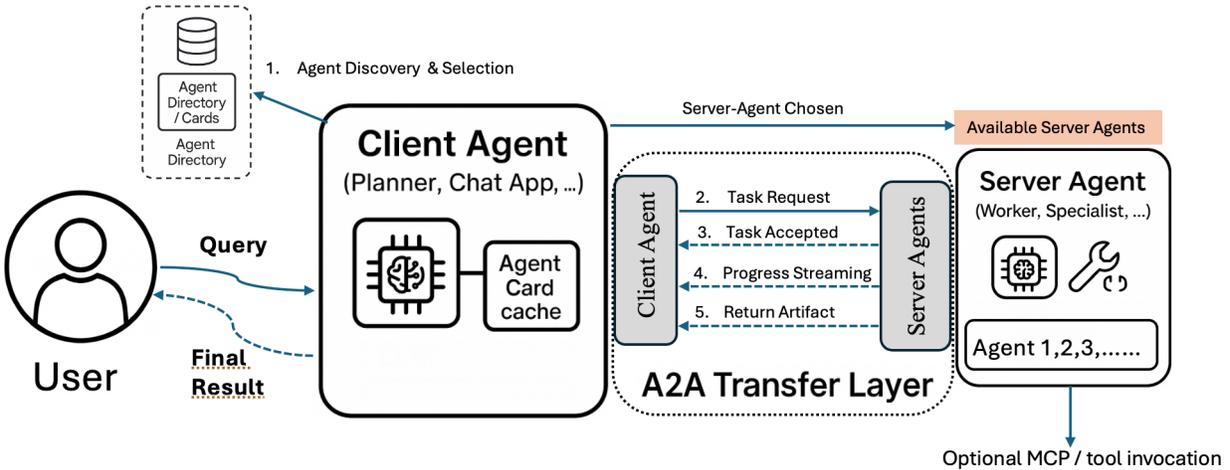

**Fig. 2.** The workflow of A2A

**A2A Example:** Consider a hiring scenario. A "Hiring Manager Agent" needs to fill a software engineer position. Using A2A, it discovers a "Recruiter Agent" via its Agent Card, which lists skills like "candidate sourcing" and supported input/output modes. The Hiring Manager Agent sends a task: "Find candidates with skill X in location Y." The Recruiter Agent accepts the task. It might use its internal capabilities (possibly via MCP) to query LinkedIn. As it finds candidates, it streams back partial results (Artifacts containing candidate profiles) via SSE over the A2A connection. The Hiring Manager Agent could then delegate subsequent tasks – scheduling interviews, background checks – to other specialized agents (Calendar Agent, HR Agent) via separate A2A tasks. A2A manages the secure communication, task state tracking, and multimodal data exchange throughout this workflow.

## 2.2 Anthropic's MCP: Standardizing Agent Context and Tool Use

The Model Context Protocol (MCP) is an open standard introduced by Anthropic that addresses how an AI model or agent accesses external data, tools, and context in a uniform way [2]. MCP defines a client–server architecture for augmenting AI models, decoupling model logic from the specifics of data sources or APIs. It acts as a "universal port" [3].

- **Core Design Principles:** MCP aims for universal connectivity (solving the "MxN" integration problem), structured context management via defined primitives (Resources, Tools), security with user control emphasis (local-first default, consent for actions), and support for two-way communication [2, 3, 8].
- **Architecture (Host/Client/Server):**
  - **Host:** The main AI application the user interacts with (e.g., Claude Desktop, IDE plugin). Manages connections and security.
  - **Client:** An intermediary within the Host, managing the connection to *one* specific MCP Server. Helps sandbox connections.
  - **Server:** A program exposing capabilities (Tools, Resources, Prompts) related to an external system (database, API, filesystem) via the MCP spec.

- **Transport & Communication:** MCP uses JSON-RPC 2.0 over two primary transports:
  - **Standard I/O (stdio):** For local client-server communication.
  - **HTTP + Server-Sent Events (SSE):** For remote connections. Client sends requests via HTTP POST; server can respond via HTTP or push asynchronous notifications/responses via SSE.
- **Discovery:** Clients discover server capabilities by invoking protocol-defined JSON-RPC methods like `tools/list`, `resources/list`, `prompts/list`. Servers can notify clients of changes (e.g., `notifications/tools/list_changed`).
- **MCP Primitives:** MCP standardizes interactions using core message types [3]:
  - **Tools:** Executable functions exposed by the server (e.g., send email, query database). Defined by a name, description, `inputSchema` (JSON Schema), and optional annotations (e.g., `destructiveHint`). Invoked via `tools/call`.
  - **Resources:** Structured data/content provided by the server for context (e.g., file content, query results). Identified by a URI, have name, description, mimeType, and content (text or base64 blob). Read via `resources/read`, updates via `resources/subscribe` and `notifications/resources/updated`.
  - **Prompts:** Predefined prompt templates or workflow scripts offered by the server, often user-selected. Defined by name, description, arguments. Used via `prompts/get` to retrieve messages for the LLM, potentially embedding resource context or defining multi-step workflows.
  - **Roots:** Client-side primitive representing host environment entry points (e.g., directories) the server might access.
  - **Sampling:** Advanced server-side primitive allowing the server to request the host AI generate text based on a prompt. Requires careful handling and often human approval due to safety implications.
- **Security:** Evolved to support OAuth 2.1, emphasizes sandboxing, user consent for tool calls, and aims to prevent prompt injection from retrieved data [12].

**MCP Example:** An AI writing assistant needs to draft an email reply incorporating calendar availability. Using MCP:

1. The assistant (Host/Client) connects to a Calendar MCP Server and an Email MCP Server.
2. To find availability, it calls `resources/read` on the Calendar server with a URI representing next week's schedule (Resource interaction).
3. To find the recipient's address, it might call `resources/read` on the Email server for the latest message thread (Resource interaction).
4. To send the email, it calls `tools/call` on the Email server with the method name "sendEmail" and parameters defined by the tool's `inputSchema` (Tool interaction). The assistant interacts via the standardized MCP interface, regardless of the underlying calendar/email providers.

## 2.3 Complementarity of A2A and MCP

A2A and MCP address distinct needs: A2A enables **horizontal** inter-agent collaboration, while MCP enables **vertical** agent-to-environment interaction. They are inherently complementary [7]. An agent might use MCP to fetch data or execute a tool, then share the results or delegate follow-up actions to another agent via A2A. Understanding this distinction is key to architecting effective integrated systems and analyzing the emergent properties of their combination.

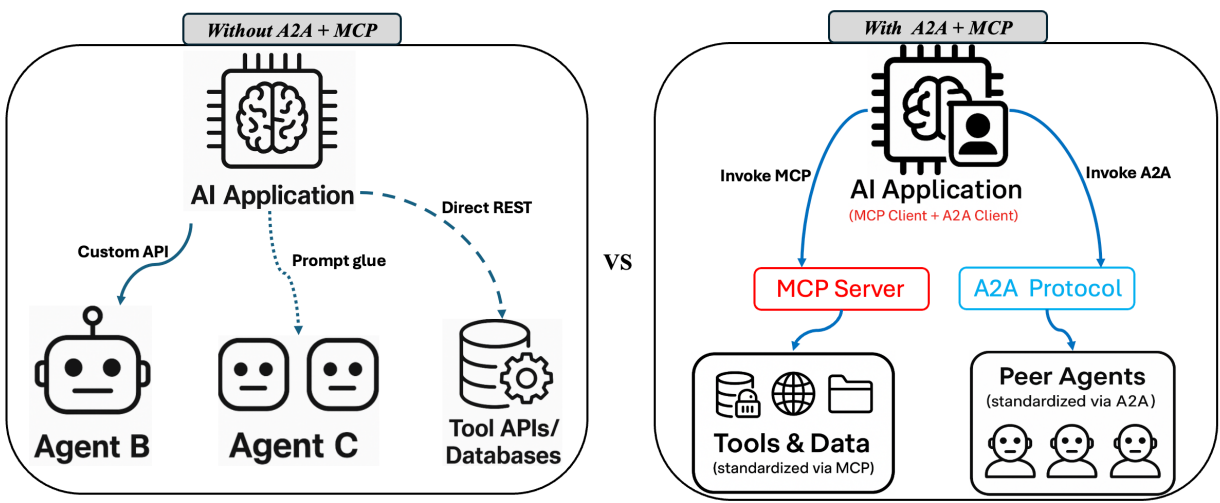

**Fig. 3.** *The workflow with and without A2A + MCP*

A comparative summary highlights their distinct roles:

| Feature | Google A2A Protocol | Anthropic MCP Protocol |
|---|---|---|
| **Primary Focus** | Agent-to-Agent communication & collaboration | Agent-to-Tool/Data Source connection & context management |
| **Integration Type** | Horizontal (Inter-Agent) | Vertical (Intra-Agent to Environment) |
| **Core Architecture** | Client Agent / Remote Agent | Host / Client / Server |
| **Key Components** | Agent Card, Task, Message, Part, Artifact | Host, Client, Server, Tool, Resource, Prompt, (Root, Sampling) |
| **Communication Protocol** | JSON-RPC 2.0 over HTTP(S), SSE, Push Notifications | JSON-RPC 2.0 over stdio, HTTP(S)+SSE |
| **Scope** | Interaction & coordination between multiple agents | Individual agent's access to external info & capabilities |
| **Intended Use Cases** | Multi-agent workflows, task delegation, cross-system collaboration | Data retrieval, tool invocation, context enrichment for agents |
| **Security Focus** | Secure agent identity, authentication, authorization, encryption | User consent, sandboxing, controlled tool/resource access |
| **Modality Support** | Text, Files, Structured Data (JSON), Forms, Audio, Video | Primarily Text, Files (Binary via blob), structured data |
| **Asynchronicity** | Native support for long-running tasks (SSE, Push) | Host manages async tool calls; server can push updates (SSE) |
| **State Management** | Explicit Task lifecycle states (submitted, working...) | Primarily request-response for Tools; Host manages workflow state |
| **Discovery Mechanism** | Agent Card (/.well-known/agent.json), Registries | Protocol methods (tools/list, resources/list, etc.) |
| **Key Strength** | Enabling complex, dynamic multi-agent collaboration | Standardizing tool/data integration & context provision |
| **Key Limitation (Initial)** | Ecosystem maturity, higher-level orchestration needed | Less emphasis on inter-agent communication, initial auth gaps |

**Table 1.** *Distinct focus and complementarity of A2A and MCP*

# 3. Taxonomy of Benefits from Integrating A2A and MCP: A Critical Evaluation

Integrating A2A and MCP offers compelling benefits, but getting them in practice requires carefully managing the complexities from combining these distinct layers.

## 3.1 Interoperability and Standardization

- **Cross-Vendor/Framework Collaboration:** Enables agents built with different tools or vendors to communicate (A2A) and access standardized tools/data (MCP), addressing a key challenge in MAS interoperability [14]. *Evaluation:* While this improves connection possibilities, achieving true *semantic* understanding—ensuring agents consistently interpret the meaning of exchanged information and capabilities—remains a significant hurdle, especially at the A2A-MCP interface [22]. This semantic gap can limit effective collaboration despite standard protocols.
- **Plug-and-Play Modular Architecture:** Allows treating agents and data connectors as swappable components [11]. *Evaluation:* This modularity is useful but needs reliable discovery methods and potentially complex coordination logic to handle dependencies. The ideal of simple "plug-and-play" might be hard to reach for complex interactions needing deep understanding between parts.
- **Consistent Governance and Compliance:** Standard protocols can simplify policy enforcement. *Evaluation:* Applying rules consistently *across* both protocols requires integrated systems that understand both agent-to-agent talks (A2A) and agent-to-tool actions (MCP). This combined governance layer isn't part of the protocols themselves and adds complexity.

## 3.2 Enhanced Capabilities via Synergy

- **Specialization and Delegation:** Creates potential for expert agent teams. A central agent (using A2A) can delegate tasks to specialized agents which use MCP for tool access. This approach is seen in many multi-agent systems. *Evaluation:* This powerful setup depends heavily on the main agent's ability to accurately match high-level A2A tasks to specific MCP tool abilities advertised by other agents. Vague skill descriptions can lead to errors. Good multi-agent planning is vital [17, 26].
- **Contextualized Reasoning with Real-Time Data:** Agents using MCP get current information, which can then be shared via A2A for better group understanding [15]. *Evaluation:* This benefit depends on efficient context sharing. Sending large amounts of raw data from MCP over A2A can be slow. Agents need good ways to summarize, check, and combine context from different MCP sources into their A2A collaboration, adding complexity.

## 3.3 Efficiency, Scalability, and Robustness

- **Parallelism and Throughput:** Enables distributing tasks across agents (A2A) for simultaneous work (using MCP). *Evaluation:* Getting significant speedups requires tasks that can truly run in parallel and smart coordination to manage dependencies and combine results, which does add its own overhead [14, 26].
- **Scalability across Domains and Load:** Makes it easier to scale by adding more agents (A2A) or tool servers (MCP). *Evaluation:* Real-world scalability depends not just on the protocols but also on factors like discovery service speed, network delays, coordination complexity, and the performance of the underlying AI models. The A2A-MCP interface itself can become a bottleneck.
- **Fault Tolerance and Graceful Failure:** Offers potential for backup options [4]. *Evaluation:* Building robust fault tolerance requires agents that can reliably detect failures (in both A2A tasks and MCP calls), assess alternatives, and replan dynamically. This makes agents much more complex than simple error handling would suggest.
- **Resource Efficiency through Delegation:** Allows using cheaper/smaller agents for simple tasks. *Evaluation:* This requires a smart coordinating agent that can accurately judge task difficulty and agent abilities (including both A2A skills and MCP tool access) to delegate efficiently – a tricky planning problem.

## 3.4 Ecosystem and Innovation Benefits

- **Network Effects and Agent Ecosystems:** Common protocols can encourage marketplaces [8]. The MCP ecosystem is growing. *Evaluation:* Success depends on wide adoption, clear standards, good discovery tools, and reliable trust/reputation systems to ensure the quality and security of third-party components – a major challenge for open systems.
- **Reduced Development Complexity:** Standard interfaces reduce *integration* code. *Evaluation:* This shifts the difficulty towards understanding the protocols, designing agent interactions *across* protocols, building effective coordination, and debugging distributed systems, requiring new skills.
- **Cross-Domain Applicability:** Protocols are general-purpose [23]. *Evaluation:* While the protocols are general, real applications need domain-specific knowledge, tools, and context. The integration itself doesn't automatically make solutions transferable between different fields.

## 3.5 Quality, Safety, and Alignment Improvements

- **Continuous Learning from Feedback:** Standardized logs provide data. *Evaluation:* Using this data effectively requires advanced methods for multi-agent learning and figuring out how to translate interaction logs into real improvements.
- **Safety via Structured Control:** Protocols offer points to insert controls. *Evaluation:* Defining and enforcing effective safety rules across dynamic A2A interactions *and* MCP tool calls is very complex. Preventing misuse, ensuring privacy, and aligning agent behavior in integrated systems are major unsolved research gaps [39].

# 4. Architectural Patterns: Integration Points and Difficulties

Integrating A2A and MCP requires architectural choices that bridge their different scopes.

1. **Pattern 1: A2A Agent Utilizing MCP Internally (Primary Pattern):** An A2A server agent uses MCP internally.
   - *Integration Insight:* This pattern keeps things clearly separated but might lead to duplicated effort if many A2A agents need the same MCP tool. Also, the A2A client doesn't directly see which MCP tools the remote agent uses, relying only on the A2A skill description, which might be vague.
2. **Pattern 2: Exposing MCP Tools via A2A Agent Card:** A2A skills directly represent MCP tools.
   - *Integration Insight:* This makes tools more discoverable via A2A but creates a **semantic mismatch**. A2A's skill format is less detailed than MCP's tool format (`inputSchema`). Trying to reliably match A2A task details to MCP tool inputs based on potentially unclear text descriptions is a major point of difficulty and potential error [1, 22].
3. **Pattern 3: A2A for Tool Orchestration (Alternative/Edge Case):** Using A2A directly for complex "tools".
   - *Integration Insight:* This uses A2A's strengths for long tasks but bypasses MCP's specific focus on standard tool interactions, potentially making tool handling less consistent across the system.

**Orchestration Layer:** Regardless of the pattern, effective integration usually needs an orchestration layer [14, 17, 26]. This layer acts as the critical hub, translating goals into A2A tasks, matching tasks to agents and their MCP abilities, managing communication, handling errors across protocols, and combining results. Designing this coordination logic, perhaps using a dedicated coordinator agent, is vital and challenging for practical A2A+MCP systems.

# 5. Economic Implications: The "Agent Economy"

The vision of an "Agent Economy" enabled by A2A+MCP integration [9] needs a critical look alongside its potential. This concept relates to the field of Agent-based Computational Economics (ACE), which models economies as dynamic systems of interacting agents [29, 30].

## 5.1 Emergence of "Agent Economy"

- **Automated Transactions:** Protocols provide the technical base. *Critical Evaluation:* A real economy requires more than just technology; it needs reliable, scalable ways to establish **trust**, manage **reputations**, handle **automated contracts**, securely **exchange value**, and

**resolve disputes** between autonomous agents. Current ideas using cryptoeconomics are still developing and have their own challenges.
- **Transformed Marketplaces:** Potential for agent-driven services/data. *Critical Evaluation:* This depends on solving **semantic discovery** (agents reliably finding and understanding services), **quality assurance** (checking agent/tool quality), and clear **liability rules** for automated transaction failures.

## 5.2 Lower Barriers and Competitive Landscape

- **Reduced Lock-in:** Open protocols *may* reduce vendor dependence. *Critical Evaluation:* Large platforms could still dominate through better coordination services, unique agent abilities, control over key data sources (via MCP), or by creating popular extensions, potentially leading to new forms of lock-in. True openness requires active community involvement.
- **Increased Choice for Buyers:** More flexibility exists. *Critical Evaluation:* This choice brings the complexity of managing parts from different vendors, ensuring compatibility beyond basic protocol rules, and potentially higher security risks from using less-vetted components.

## 5.3 New Roles and Business Models

- **Agent Brokers/Orchestrators, MCP Server Providers, Agent Auditors:** These roles seem possible [5, 13, 20]. *Critical Evaluation:* Their success depends on proving their value compared to integrated solutions. Auditing complex AI agents that use both A2A and MCP requires new methods.

## 5.4 Productivity and Cost Impacts

- **Productivity Gains:** Automation potential is clear [23, 24]. *Critical Evaluation:* Real gains must outweigh the significant costs of building, securing, managing, and debugging these complex systems. Automation benefits might initially be limited to specific, well-defined areas.
- **Workforce Transformation:** Job changes are likely. *Critical Evaluation:* Society needs to adapt proactively. There's a risk of increasing inequality if benefits mainly go to owners or specialized AI developers.
- **Efficiency Paradox:** Automation might increase overall resource use. *Critical Evaluation:* Optimizing energy and computing efficiency for large agent systems will be important for sustainability.

## 5.5 Strategic Control and Data Value

- **Protocol Stewardship:** Influence follows standards. *Critical Evaluation:* Keeping protocols truly open requires active community participation to prevent control by a few dominant companies.
- **Data Value & Licensing:** MCP makes data more usable. *Critical Evaluation:* Standardized

access via MCP raises complex issues around data privacy, ownership, fair payment for data use, and potential information advantages for agents with special data access, especially if this data influences A2A interactions [18].

# 6. Risks and Challenges at the A2A-MCP Intersection

Integrating A2A and MCP creates unique challenges beyond those inherent in each protocol individually.

## 6.1 Key Risks and Challenges

1. **Security - Compounded Vulnerabilities:** The combination creates novel attack vectors.
   - *Integration-Specific Threats:* A key risk involves attackers using A2A discovery to find agents, then exploiting vulnerabilities exposed through those agents' MCP tool connections [19, 21]. For example, finding an agent via A2A that uses an insecure MCP server is an indirect attack path. Threats like "tool squatting" (maliciously registering fake tools) are more dangerous when combined with A2A's discovery potential [32]. Malicious instructions can also spread between agents via A2A, potentially triggering unsafe MCP tool actions [28, 35, 37].
   - *Identity and Trust Management:* Establishing trust is harder. An agent must trust its A2A partner and, indirectly, the tools that partner uses via MCP. Managing security across both A2A talks and later MCP calls requires complex solutions not built into the basic protocols. Reliable trust and reputation models are crucial but difficult to implement [20, 36].
2. **Semantic Interoperability Gap:** Ensuring agents understand each other *and* the tools they access is critical but difficult.
   - *Integration-Specific Challenge:* A major difficulty is translating high-level A2A tasks into specific, correct MCP tool commands. An A2A client might ask for a summary, but the remote agent needs to figure out the exact MCP queries and tool parameters. A2A's skill descriptions often lack the detail of MCP's tool requirements, leading to potential errors at the integration point [1, 22]. This needs better ways for agents to negotiate meaning or use shared knowledge bases.
3. **Debugging and Observability Complexity:** Diagnosing failures in integrated systems is significantly harder.
   - *Integration-Specific Challenge:* If a workflow that uses multiple agents and tools breaks down, diagnosing the problem involves tracking activities across A2A tasks, agent communications, and MCP tool interactions. To pinpoint the error (like an A2A misunderstanding, a faulty MCP tool, or bad mapping), we need integrated monitoring tools that cover both protocols, but such tools don't exist yet. Formal verification methods show promise but struggle with the scale of these combined systems [25, 31].

4. **Orchestration Complexity:** Designing the coordinating logic that manages the flow across A2A and MCP is challenging. The orchestrator must handle planning, delegation, monitoring, error handling across both protocols, and combining results [14, 17, 26].
5. **Governance and Policy Enforcement:** Defining and enforcing consistent security, privacy, and ethical rules across both agent-to-agent (A2A) and agent-to-tool (MCP) interactions requires a unified approach.

## 6.2 Future Directions

Addressing the integration challenges requires focused research and development:

1. **Semantic Negotiation and Shared Ontologies:** Developing ways for agents to clarify task meanings, capabilities beyond simple Agent Cards, and reliably map A2A requests to MCP tool details [22]. Shared knowledge structures (ontologies) could help.
2. **Integrated Security Frameworks:** Designing security systems that handle the combined A2A+MCP risks, including identity management, cross-protocol permissions, and defenses against threats like tool squatting [21, 28, 32, 38]. Improving MCP's built-in security features is also needed.
3. **Cross-Protocol Observability and Debugging Tools:** Creating tools for unified logging, tracing, and visualizing activity across both A2A and MCP to make debugging easier.
4. **Robust Multi-Agent Orchestration and Planning:** Improving multi-agent planning techniques [14, 17, 26] specifically for coordinating agents that use external tools via MCP.
5. **Formal Verification for Integrated Systems:** Adapting formal methods to verify the correctness and safety of combined A2A+MCP systems [25, 31].
6. **Trust and Reputation Infrastructure:** Building reliable trust and reputation systems for open ecosystems involving both A2A communication and MCP tool use.
7. **Standardized Governance Mechanisms:** Developing ways to define and enforce operational, security, and ethical rules consistently across integrated A2A+MCP systems, especially for any future "Agent Economy" [13].

## 6.3 Outlook

The integration of A2A and MCP provides a powerful, standardized foundation. However, near-term success depends on the community building good tools, establishing best practices for managing integration complexity (especially security and semantics), and creating a trustworthy ecosystem. Long-term adoption requires solving the deeper challenges of orchestration, governance, and verification for these complex systems.

# 7. Conclusion

The combination of Google's A2A and Anthropic's MCP protocols marks a significant step towards enabling more sophisticated, interoperable, and scalable multi-agent systems. By standardizing horizontal (agent-to-agent) and vertical (agent-to-environment) interactions, respectively, they offer a pathway to move beyond brittle, custom integrations. This paper has provided a critical analysis of this integration, acknowledging the substantial benefits while focusing on the unique challenges and complexities that emerge specifically at the intersection of these two standards.

Our analysis reveals that while the potential for enhanced capabilities, efficiency, and ecosystem growth is considerable, realizing these benefits requires confronting significant hurdles. Key challenges include ensuring semantic alignment between A2A tasks and MCP tools, managing compounded security risks from agent discovery and tool execution, the difficulty of debugging across both protocols, and establishing governance for complex interactions, particularly for the "Agent Economy." Special attention shall also be paid to specific security risks associated with the MCP server lifecycle.

Ultimately, A2A and MCP provide crucial architectural building blocks. However, they are not a complete solution. Unlocking the full potential of collaborative, tool-using AI agents requires continued innovation in multi-agent planning and orchestration, robust trust and reputation mechanisms, advanced security architectures tailored for integrated systems, practical solutions for semantic negotiation, and effective governance models. Addressing these integration-specific challenges will be essential for the successful development and deployment of the next generation of truly capable and reliable multi-agent AI systems.